
\magnification=1200
\baselineskip=20pt
\def\lsim{<\kern-2.5ex\lower0.85ex\hbox{$\sim$}\ }
\def\rsim{>\kern-2.5ex\lower0.85ex\hbox{$\sim$}\ }
\overfullrule=0pt

\tolerance 10000
\centerline{ }
\vskip 1cm
\hfill UR 1275
\noindent

\hfill ER-40685-727
\bigskip
\centerline {\bf FEYNMAN PARAMETRIZATION AND THE}
\centerline {\bf DEGENERATE ELECTRON GAS}
\vskip 2cm
\centerline{ Paulo F. Bedaque}
\centerline{ and}
\centerline{ Ashok Das}
\bigskip
\centerline{\it Department of Physics and Astronomy}
\centerline{\it University of Rochester}
\centerline{\it Rochester, NY 14627, USA}
\vskip 1cm
\centerline{\bf Abstract}
\bigskip
We give an alternate derivation of Weldon's formula for combining products
of factors with non identical analytic behavior. While such a formula would
appear to be useful in finite
 temperature calculations, we give an example of a
zero temperature calculation,
 namely, the degenerate electron gas, to justify the result.
\vfill
\eject
{\bf I. Introduction}
\bigskip
The question of analyticity in finite temperature Green's functions has led to
some disagreement between imaginary time and real time calculations in the
recent years. To be specific, the
 real part of the self-energy of a scalar theory was shown [1], in
the imaginary time formalism [2],
 to have a non analytic behaviour at vanishing
external momentum, namely, at $p^\mu = 0$. A modification of the
imaginary time calculation was proposed in [3]
which used Feynman parametrization
 and led to a self-energy which is analytic at
$p^\mu\rightarrow 0$. The real time [4] calculation
 of the self-energy was carried out with a limiting procedure [5]
 leading to a result which coincided with that of ref. [1]
 whereas a calculation done with
$\epsilon$-regularization [6] led to a self-energy which is analytic at
$p^\mu=0$.
 It is worth pointing out here that the real time
calculation in ref. [6]
 used the naive Feynman parametrization formula in combining
products of factors. The disagreement between these various calculations was
recently resolved by Weldon
 [7] who showed that in combining factors of opposite
analytic behavior, the naive Feynman parametrization needs to be modified.
At finite temperature the factors one needs to combine often have opposite
analytic behavior and with the modification of the combination formula, Weldon
showed that all calculations agree.

In a separate development, we have been interested in studying the degenerate
electron gas, for which many results are known at zero temperature [8,9],
 using the
$\epsilon$-regularization. Our primary interest in this system
  arose mainly
because the real part of the self-energy in this system
 also shows a nonanalytic
behavior at $p^\mu = 0$ even at zero temperature.
Surprisingly enough our calculation using the
$\epsilon$-regularization disagreed with the standard classical value.
The $\epsilon$-regularization, therefore, appeared suspect. However, as we will
show shortly, the $\epsilon$-regularization itself is well behaved. But this is
one of the few cases at zero temperature where the Feynman formula needs to be
modified according to Weldon's formula.
 The plan of the paper is as follows. In sec. II, we give an alternate
derivation of Weldon's formula in a more general form. In sec. III, we
calculate the real part of the
self-energy of a degenerate electron gas at zero temperature
by integrating out the $\delta$-function directly. In sec. IV, we calculate the
same quantity using
 the $\epsilon$-regularization but without using Feynman
parametrization and show that the two results are the same. In sec. V the same
calculation is carried out using the $\epsilon$-regularization but with the
naive Feynman parametrization formula. The result in this case is
different from the earlier ones
 showing that the combination formula is the real culprit.
In sec. VI, we use the modified combination formula to show that the result
agrees with the earlier calculations. Finally, we present our conclusions in
sec. VII and comment on attempts to modify Feynman rules at finite
temperature.
\bigskip
\bigskip
{\bf II. Alternate Derivation of Weldon's Formula}
\bigskip
Let us consider the integral
$$ \eqalignno{I=& P\int^1_0 {dx\over [x(A+i\alpha\epsilon)+(1-x)
(B+i\beta\epsilon)]^2}\cr
=&P\int^1_0 {dx\over [x(A-B+i\epsilon(\alpha-\beta))+B+i\beta\epsilon]^2} ,
&(1)\cr}$$

\noindent
where $A, B, \alpha, \beta$ and $\epsilon$ are assumed to be real. Furthermore,
we assume that $\epsilon\rightarrow 0^+$ and that $\alpha , \beta$ have unit
 magnitude (i.e., $\alpha=\pm 1, \beta=\pm 1$).

If there are no poles of the integrand on the real $x$-axis
between ($0,1$) then we
identify

$$\eqalignno{I=&P\int^1_0 {dx\over
[x(A-B+i\epsilon(\alpha-\beta))+B+i\beta\epsilon]^2}\cr
=&\int^1_0 {dx\over [x(A-B+i\epsilon(\alpha-\beta))+B+i\beta\epsilon]^2}\cr
=& -{1\over A-B+i(\alpha-\beta)\epsilon}\quad
{1\over x(A-B+i(\alpha-\beta)\epsilon+b+i\beta\epsilon}
\bigg|^1_0\cr
=&{1\over A+i\alpha\epsilon}\quad {1\over B+i\beta\epsilon} .&(2)\cr}$$
\noindent
This is, of course, the naive Feynman parametrization formula and we note that
since $\alpha$ and $\beta$ can
 always be rescaled, we have lost no generality in
choosing their magnitudes to be unity.

If, on the other hand, the integrand has poles on the real $x$-axis between
($0,1$), then we have to regularize the integrand by taking its principal
value.
Note that the integrand will have a pole at $x=x_0$ on the real axis if
$$ x_0(A-B+i(\alpha-\beta)\epsilon)+B+i\beta\epsilon=0\eqno(3)$$
\noindent

$${\rm or, }
\quad x_0(A-B)+B= -i(\beta+x_0(\alpha-\beta))\epsilon=0 .\eqno(4)$$
\noindent
The solution to these equations are
$$ x_0={\beta\over \beta-\alpha}\eqno(5)$$
$$ \beta A = \alpha B .\eqno(6)$$
\noindent
It is clear now from Eq.
 (5) that $0\le x_0\le 1$ only when $\alpha$ and $\beta$
have opposite sign (if $\alpha =1, \beta=-1$, then $x_0={1\over 2}$ but we will
leave $\alpha , \beta$ arbitrary at this point).

Assuming that $0\le x_0 = {\beta\over \beta-\alpha} \le 1$, we can evaluate the
integral in Eq. (1) as
$$\eqalignno{I= &\, \lim_{\eta\rightarrow 0^+}
\bigg[ \int_0^{{\beta\over \beta-\alpha}
-\eta}{dx\over [x(A-B+i\epsilon(\alpha-\beta))+B+i\beta\epsilon]^2}\cr
&\qquad\qquad +\int^1_{{\beta\over \beta-\alpha}+\eta}
{dx\over [x(A-B+i\epsilon(\alpha-\beta))+B+i\beta\epsilon]^2}\bigg]\cr
=&\,{1\over A-B+i(\alpha-\beta)\epsilon}\quad \lim_{\eta\rightarrow 0^+}
\big[ -{1\over x(A-B+i\epsilon(\alpha-\beta))+B+i\beta\epsilon}
\bigg|_0^{{\beta\over \beta-\alpha}-\eta}\cr
&\phantom {{1\over A-B+i(\alpha-\beta)\epsilon}\quad
 \lim_{\eta\rightarrow 0^+}}
-{1\over x(A-B+i\epsilon(\alpha-\beta))+B+i\beta\epsilon}
\bigg|^1_{{\beta\over \beta-\alpha}
+\eta}\Bigr]\cr
=&\,{1\over A-B+i(\alpha-\beta)\epsilon}\quad ({1\over B+i\beta\epsilon}
-{1\over A+i\alpha\epsilon})\cr
&-{1\over A-B+i(\alpha-\beta)\epsilon}\quad \lim_{\eta\rightarrow 0^+}
\bigl [{\beta-\alpha\over \beta A-\alpha B+i\eta\epsilon}-
{\beta-\alpha\over \beta A-\alpha B-i\eta\epsilon}\bigr ]\cr
=&\,{1\over A+i\alpha\epsilon}{1\over B+i\beta\epsilon}\cr
-&{\beta-\alpha\over A-B+i(\alpha-\beta)\epsilon}
\bigl[ P {1\over \beta A-\alpha B}-i\pi\delta(\beta A-\alpha B)
- P {1\over \beta A-\alpha B}-i
\pi\delta(\beta A-\alpha B)\bigr]\cr
=\  &{1\over A+i\alpha\epsilon}{1\over B+i\beta\epsilon}
-2\pi i {(\alpha-\beta)\delta(\beta A -\alpha B)\over
A-B+i(\alpha-\beta)\epsilon} . &(7)\cr}$$
Comparing Eqs. (2) and (7) we note that for any value of $\alpha$ and $\beta$
(with unit magnitude), we can write
$$\eqalignno{{1\over A+i\alpha\epsilon}{1\over B+i\beta\epsilon}=&I+
2\pi i {(\alpha-\beta)\delta(\beta A -\alpha B)\over
A-B+i(\alpha-\beta)\epsilon}\cr
=&P\int^1_0 {dx\over [x(A+i\alpha\epsilon)+(1-x)
(B+i\beta\epsilon)]^2}+2\pi i {(\alpha-\beta)\delta(\beta A -\alpha B)\over
A-B+i(\alpha-\beta)\epsilon} .&(8)\cr}$$
\noindent
This clearly reduces to Weldon's formula (Eq. (18) in ref. [7]) when $\alpha=1$
and $\beta=-1$. This is the modified Feynman parametrization formula.
\bigskip
\bigskip
{\bf III. Degenerate Electron Gas (Standard Results):}
\bigskip
Let us consider a nonrelativistic, degenerate electron gas at $T=0$ and $\mu\ne
0$ [8,9]
 where the chemical potential $\mu$ can be identified with the Fermi energy.
In this case, the propagator has the form
$$\eqalignno{S(p)=&\,{i\over p_0-\omega_p+\mu+i\epsilon \,{\rm sgn}
(\omega_p-\mu)}\cr
=&\,{i\over p_0-\omega_p+\mu+i\epsilon \,{\rm sgn}(p_0)} .&(9)\cr}$$
\noindent
with $\omega_p = {\vec p^2 \over 2m}$.
We would like to emphasize here that while the two forms of the propagator in
Eq. (9) are equivalent, it is the second form that is easier to use in
practical
calculations. We would also like to note here that the propagator in this case
does not have a unique analytic behaviour in the sense that the location of the
singularity depends on the sign of $p_0$. This is quite reminiscent of the non
unique analytic behavior of
propagators at finite temperature even though the
system under study is at $T=0$.

We can calculate the real part of the self-energy in a standar manner.
$$\eqalignno{Re\, \pi(p)=& \,Re \bigl[-i\int{d^4k\over (2\pi)^4}
S(k) S(k-p)\bigr ]\cr
=&\,Re\, [-i\int{d^4k\over (2\pi)^4}
({i\over k_0-\omega_k+\mu+i\epsilon\, {\rm sgn}(k_0)})\cr
&\phantom {Re\, [-i\int{d^4k\over (2\pi)^4}}\times
({i\over k_0-p_0-\omega_{k-p}+\mu+i\epsilon\, {\rm sgn}(k_0-p_0)})\cr
=&\,{1\over 2}\int{d^4k\over (2\pi)^3}\,[P{1\over k_0-\omega_k+\mu}
\,{\rm sgn}(k_0-p_0)\delta(k_0-p_0-\omega_{k-p}+\mu)\cr
&\qquad\qquad+P{1\over k_0-p_0-\omega_{k-p}+\mu}
\,{\rm sgn}(k_0)\delta(k_0-\omega_k+\mu)]\cr
=&\,{1\over 2}\int{d^4k\over (2\pi)^3}\,[P{1\over k_0+p_0-\omega_{k+p}+\mu}
+p_\mu\rightarrow -p_\mu]\,{\rm sgn}(k_0)\delta(k_0-\omega_k+\mu) .&(10)\cr}$$
\noindent
Here we have let $k_\mu\rightarrow k_\mu+p_\mu$ in the first term
 of the integrand to arrive
at the form in the last step. We have also used the standard result
$$\lim_{\epsilon\rightarrow 0^+}{1\over x+i\alpha\epsilon}
\,=\,P{1\over x}-i\pi \, {\rm sgn}(\alpha)\delta(x)\, .\eqno(11)$$
We can do the $k_0$ integration in Eq. (10)
directly using the delta function to obtain
$$Re\ \,\pi(p)={1\over 2}\int{d^3k\over (2\pi)^3}[P{1\over
p_0+\omega_k-\omega_{k+p}}+ p_\mu\rightarrow -p_\mu]\, {\rm sgn}
(\omega_k-\mu) .\eqno(12)$$
\noindent
This is the standard result and can be rewritten in a straightforward manner to
agree with the form of the result in ref. [9].
\bigskip
\bigskip
{\bf IV. $\epsilon$-Regularization ( Without Feynman Parametrization):}
\bigskip
Let us next use the $\epsilon$-regularization, namely,
$$ P{1\over x}= \lim_{\epsilon\rightarrow 0^+}\,{x\over x^2+\epsilon^2}=
\lim_{\epsilon\rightarrow 0^+}\,{1\over 2}\bigl ( {1\over x+i\epsilon}+
{1\over x-i\epsilon}\bigr )\eqno(13)$$
$$\delta(x)=\lim_{\epsilon\rightarrow 0^+}\,{1\over \pi}{\epsilon\over
x^2+\epsilon^2}=\lim_{\epsilon\rightarrow 0^+}\,{i\over 2\pi}
\bigl ( {1\over x+i\epsilon}-{1\over x-i\epsilon}\bigr )\eqno(14)$$
\noindent
In this case, the relation in Eq. (10) can be written
as (From now on we will not
explicitly write $\epsilon\rightarrow 0^+$ although this should be understood.)
$$\eqalign{Re  \,\pi (p)=
{i\over 4}\int{d^4k\over (2\pi)^4}\bigl [&
{1\over k_0+p_0-\omega_{k+p}+\mu+i\epsilon}+
{1\over k_0+p_0-\omega_{k+p}+\mu-i\epsilon}\cr
&+p_\mu\rightarrow -p_\mu\bigr ]\,
 {\rm  sgn}(k_0)\big
({1\over k_0-\omega_k+\mu+i\epsilon}-{1\over k_0-\omega_k+\mu-i\epsilon}
\bigr )\cr}$$
$$\eqalignno{\phantom{Re\,\pi(p)}={i\over 4}\int{d^4k\over (2\pi)^4}
\,{\rm sgn}(k_0)&\bigl [
{1\over k_0+p_0-\omega_{k+p}+\mu+i\epsilon}\,{1\over
k_0-\omega_k+\mu+i\epsilon}\cr
&-{1\over k_0+p_0-\omega_{k+p}+\mu-i\epsilon}\,{1\over
k_0-\omega_k+\mu-i\epsilon}\cr
&+{1\over k_0+p_0-\omega_{k+p}+\mu-i\epsilon}\,{1\over
k_0-\omega_k+\mu+i\epsilon}\cr
&-{1\over k_0+p_0-\omega_{k+p}+\mu+i\epsilon}\,{1\over
k_0-\omega_k+\mu-i\epsilon}\cr
&+p_\mu\rightarrow -p_\mu\bigr ]\cr
&=I_1+I_2 \, ,&(15)\cr}$$
\noindent
where we have separated the integral into two terms.
 $I_1$ contains products of terms with similar analytic behavior while the
factors in $I_2$ have opposite analytic behavior. We note that
$$\eqalign{I_1={i\over 4}\int{d^4k\over (2\pi)^4}
\,{\rm sgn}(k_0)\bigl [&
{1\over k_0+p_0-\omega_{k+p}+\mu+i\epsilon}\, {1\over
k_0-\omega_k+\mu+i\epsilon}\cr
&-{1\over k_0+p_0-\omega_{k+p}+\mu-i\epsilon}\, {1\over
k_0-\omega_k+\mu-i\epsilon}\cr
&+p_\mu\rightarrow -p_\mu\bigr ]\cr}$$
$$\eqalignno{\phantom {I_1}={i\over 4}\int{d^3k\over (2\pi)^4}
\int_0^\infty dk_0\bigl [&{1\over k_0+p_0-\omega_{k+p}+\mu+i\epsilon}{1\over
k_0-\omega_k+\mu+i\epsilon}\cr
&-{1\over k_0-p_0+\omega_{k+p}-\mu-i\epsilon}{1\over
k_0+\omega_k-\mu-i\epsilon}\cr
&-{1\over k_0+p_0-\omega_{k+p}+\mu-i\epsilon}{1\over
k_0-\omega_k+\mu-i\epsilon}\cr
&+{1\over k_0-p_0+\omega_{k+p}-\mu+i\epsilon}{1\over
k_0+\omega_k-\mu+i\epsilon}\cr
&+p_\mu\rightarrow -p_\mu\bigr ] .&(16)\cr}$$
\noindent
This can be evaluated with the method of the residues using a contour in the
upper right quadrant and the result is
$$\eqalignno{I_1={i\over 4}\int{d^3k\over (2\pi)^4} 2\pi i \bigl [&
-{\theta(\mu+p_0-\omega_{k+p})\over p_0+\omega_k-\omega_{k+p}}
-{\theta(\mu-\omega_k)\over \omega_{k+p}-p_0-\omega_k}\cr
&-{\theta(-\mu-p_0+\omega_{k+p})\over \omega_{k+p}-p_0-\omega_k}
-{\theta(-\mu+\omega_k)\over \omega_k-\omega_{k+p}+p_0}\cr
&+p_\mu\rightarrow -p_\mu\bigr ] \cr
={1\over 4}\int{d^3k\over (2\pi)^3}\,\,\bigl [
&{1\over p_0+\omega_k-\omega_{k+p}}
\bigl({\rm sgn}(\omega_k-\mu)-{\rm sgn }(\omega_{k+p}-p_0-\mu)\bigr)\cr
&+p_\mu\rightarrow -p_\mu\bigr ] .&(17)\cr}$$
\noindent
Similarly, we can calculate
$$\eqalign{I_2={i\over 4}\int{d^4k\over (2\pi)^4}
\,{\rm sgn}(k_0)\bigl [&
{1\over k_0+p_0-\omega_{k+p}+\mu-i\epsilon}{1\over
k_0-\omega_k+\mu+i\epsilon}\cr
-&{1\over k_0+p_0-\omega_{k+p}+\mu+i\epsilon}{1\over
k_0-\omega_k+\mu-i\epsilon}\cr
+&p_\mu\rightarrow -p_\mu\bigr ]\cr}$$
$$\eqalignno{\phantom {I_2}={i\over 4}\int{d^3k\over (2\pi)^4}
\int_0^\infty dk_0\bigl [&{1\over k_0+p_0-\omega_{k+p}+\mu-i\epsilon}{1\over
k_0-\omega_k+\mu+i\epsilon}\cr
-&{1\over k_0-p_0+\omega_{k+p}-\mu+i\epsilon}{1\over
k_0+\omega_k-\mu-i\epsilon}\cr
-&{1\over k_0+p_0-\omega_{k+p}+\mu+i\epsilon}{1\over
k_0-\omega_k+\mu-i\epsilon}\cr
+&{1\over k_0-p_0+\omega_{k+p}-\mu-i\epsilon}{1\over
k_0+\omega_k-\mu+i\epsilon}\cr
+&p_\mu\rightarrow -p_\mu\bigr ] &(18)\cr}$$
\noindent
which, by the method of residues has the value
$$\eqalignno{I_2={i\over 4}\int{d^3k\over (2\pi)^4} \,2\pi i \bigl [&
{\theta(-\mu-p_0+\omega_{k+p})\over \omega_{k+p}-p_0-\omega_k+2i\epsilon}
-{\theta(\mu-\omega_k)\over \omega_{k+p}-p_0-\omega_k+2i\epsilon}\cr
&+{\theta(\mu+p_0-\omega_{k+p})\over p_0+\omega_k-\omega_{k+p}+2i\epsilon}
-{\theta(\omega_k-\mu)\over \omega_k-\omega_{k+p}+p_0+2i\epsilon}\cr
&+p_\mu\rightarrow -p_\mu\bigr ] .&(19)\cr}$$
\noindent
Using Eq. (11) we can simplify this to
$$\eqalign{I_2={1\over 4}\int{d^3k\over (2\pi)^3}\Bigl [ &
{1\over p_0+\omega_k-\omega_{k+p}}
\bigl({\rm sgn}(\omega_k-\mu)+{\rm sgn}(\omega_{k+p}-p_0-\mu)\bigr)
+p_\mu\rightarrow -p_\mu\Bigr ] .\cr}\eqno(20)$$
\noindent
Adding the two contributions in Eqs. (17) and (20), we obtain
$$\eqalignno{Re \pi (p)=&I_1+I_2\cr
=&{1\over 2}\int{d^3k\over (2\pi)^3}\bigl [
{1\over p_0+\omega_k-\omega_{k+p}}
+p_\mu\rightarrow -p_\mu\bigr ]\, {\rm sgn}(\omega_k-\mu) .&(21)\cr}$$
\noindent
This is, of course, the same result as in Eq. (12) obtained by a direct
calculation using the $\delta$ function. This calculation shows that the
$\epsilon$-regularization is well defined and is free from any difficulty.
\bigskip
\bigskip
{\bf V. $\epsilon$-Regularization (With naive Feynman Parametrization):}
\bigskip
If we naively use the Feynman parametrization
$${1\over A+i\alpha\epsilon}{1\over B+i\beta\epsilon}=
\int^1_0 {dx\over [x(A+i\alpha\epsilon)+(1-x)
(B+i\beta\epsilon)]^2}\eqno(22)$$
\noindent
then we can write (see Eq. (16))
$$\eqalign{I_1={i\over 4}\int{d^3k\over (2\pi)^4}
\int_0^\infty dk_0\int _0^1 dx
\Bigl [&{1\over [k_0+x(p_0-\omega_{k+p})-(1-x)\omega_k+\mu+i\epsilon]^2}\cr
-&{1\over [k_0-x(p_0-\omega_{k+p})+(1-x)\omega_k-\mu-i\epsilon]^2}\cr
-&{1\over [k_0+x(p_0-\omega_{k+p})-(1-x)\omega_k+\mu-i\epsilon]^2}\cr
+&{1\over [k_0-x(p_0-\omega_{k+p})+(1-x)\omega_k-\mu+i\epsilon]^2}\cr
+&p_\mu\rightarrow -p_\mu\Bigr ]\cr}$$

$$\eqalignno{\phantom{I_1}
={i\over 4}\int{d^3k\over (2\pi)^4} \int _0^1 dx
 &\,\Bigl[{1\over x(p_0+\omega_k-\omega_{k+p})-\omega_k+\mu+i\epsilon}\cr
+&{1\over x(p_0+\omega_k-\omega_{k+p})-\omega_k+\mu+i\epsilon}\cr
-&{1\over x(p_0+\omega_k-\omega_{k+p})-\omega_k+\mu-i\epsilon}\cr
-&{1\over x(p_0+\omega_k-\omega_{k+p})-\omega_k+\mu-i\epsilon}\cr
+&p_\mu\rightarrow -p_\mu \Bigr] &(23  )\cr}$$
\noindent
This can be simplified using Eq. (11) to obtain

$$\eqalignno{I_1&={i\over 2}
\int{d^3k\over (2\pi)^4}\int _0^1 dx\bigl [
-2\pi i \delta(x(p_0+\omega_k-\omega_{k+p})-\omega_k+\mu)
+p_\mu\rightarrow -p_\mu\bigr ]\cr
&={1\over 2}\int{d^3k\over (2\pi)^3}\bigl [
\int_{-\omega_k+\mu}^{p_0+\mu-\omega_{k+p}}ds{\delta(s)\over p_0+\omega_k
-\omega_{k+p}}+p_\mu\rightarrow -p_\mu\bigr ]\cr
&={1\over 4}\int{d^3k\over (2\pi)^3}\bigl [{1\over
p_0+\omega_k-\omega_{k+p}} ({\rm sgn}(\omega_k-\mu)
-{\rm sgn}(\omega_{k+p}-p_0-\mu))
+p_\mu\rightarrow -p_\mu\bigr ] .&(24)\cr}$$
\noindent
We see that this is exactly the same as Eq. (17) which shows that the naive
Feynman parametrization holds for factors with similar analytic behavior.

Using Eq. (18), we can calculate $I_2$ and it has the form
$$\eqalign{I_2={i\over 4}\int{d^3k\over (2\pi)^4}\int_0^\infty dk_0\int _0^1 dx
\bigl [&{1\over
[k_0+x(p_0-\omega_{k+p})-(1-x)\omega_k+\mu+i(1-2x)\epsilon]^2}\cr
-&{1\over [k_0-x(p_0-\omega_{k+p})+(1-x)\omega_k-\mu-i(1-2x)\epsilon]^2}\cr
-&{1\over [k_0+x(p_0-\omega_{k+p})-(1-x)\omega_k+\mu-i(1-2x)\epsilon]^2}\cr
+&{1\over [k_0-x(p_0-\omega_{k+p})+(1-x)\omega_k-\mu+i(1-2x)\epsilon]^2}\cr
+&p_\mu\rightarrow -p_\mu\bigr ]\cr}$$
$$\eqalignno{\phantom{I_2}={i\over 4}\int{d^3k\over (2\pi)^4}\int _0^1 dx
\bigl[ &{1\over x(p_0+\omega_k-\omega_{k+p})-\omega_k+\mu+i(1-2x)\epsilon}\cr
+&{1\over x(p_0+\omega_k-\omega_{k+p})-\omega_k+\mu+i(1-2x)\epsilon}\cr
-&{1\over x(p_0+\omega_k-\omega_{k+p})-\omega_k+\mu-i(1-2x)\epsilon}\cr
-&{1\over x(p_0+\omega_k-\omega_{k+p})-\omega_k+\mu-i(1-2x)\epsilon}\cr
+&p_\mu\rightarrow -p_\mu\bigr ]\cr
={1\over 2}\phantom{[}\int{d^3k\over (2\pi)^3}\int _0^1 dx\bigl [
&{\rm sgn}(1-2x)\delta(x(p_0+\omega_k-\omega_{k+p})-\omega_k+\mu)
+p_\mu\rightarrow -p_\mu\bigr ]\cr
={1\over 2}\int{d^3k\over (2\pi)^3}\biggl [ \int_0^{1\over 2} dx\bigl [
&\delta(x(p_0+\omega_k-\omega_{k+p})-\omega_k+\mu)
+p_\mu\rightarrow -p_\mu\bigr ]\cr
-\int^1_{1\over 2} dx\bigl [
&\delta(x(p_0+\omega_k-\omega_{k+p})-\omega_k+\mu)
+p_\mu\rightarrow -p_\mu\bigr ]\biggr ] .&(25)\cr}$$
\noindent
This can be evaluated as in Eq. (24) and we obtain
$$\eqalignno{I_2={1\over 4}\int{d^3k\over (2\pi)^3}\biggl [
{1\over p_0+\omega_k-\omega_{k+p}}\bigl [
&{\rm sgn}(\omega_k-\mu)+{\rm sgn}(\omega_{k+p}-p_0-\mu)\cr
&+2\,{\rm sgn}({p_0-\omega_k-\omega_{k+p}
\over 2}+\mu)\bigr ]\cr
&+p_\mu\rightarrow -p_\mu\biggr ] .&(26)\cr}$$

We notice that $I_2$ calculated with the naive Feynman parametrization
differs from Eq. (20) and, therefore,
 it is clear that $Re \,\pi(p)$ which is the sum of
$I_1$ and $I_2$ will be different in this case from the standard result. The
reson for the discrepancy is also clear. The $\epsilon$-regularization has no
problems, however, it is the Feynman combination formula in the case of factors
with opposite analytic behavior which seems to be suspect.
\bigskip
\bigskip
{\bf VI. $\epsilon$-Regularization ( Weldon's correction ):}
\bigskip
As we see from Eq. (8), Weldon's corrections apply only when we are combining
factors with opposite analytic behaviour. This is, of course, the case for
$I_2$. Thus, with Eq. (8) we note that $I_2$ will pick up an additional
contribution of the form
$$\leftline{\indent$\displaystyle\eqalign{\tilde I_2={i\over 4}
\int {d^3k\over (2\pi)^4}\int_0^\infty dk_0 \bigl [&
-{4\pi i \,\delta(2k_0+p_0-\omega_k-\omega_{k+p}+2\mu)\over
p_0+\omega_k-\omega_{k+p}-2i\epsilon}\cr
&+{4\pi i \,\delta(2k_0-p_0+\omega_k+\omega_{k+p}-2\mu)\over
p_0+\omega_k-\omega_{k+p}-2i\epsilon}\cr
&-{4\pi i \,\delta(2k_0+p_0-\omega_k-\omega_{k+p}+2\mu)\over
p_0+\omega_k-\omega_{k+p}+2i\epsilon}\cr
&+{4\pi i \,\delta(2k_0-p_0+\omega_k+\omega_{k+p}-2\mu)\over
p_0+\omega_k-\omega_{k+p}+2i\epsilon}\cr
&+p_\mu\rightarrow -p_\mu\bigr] \cr}$}$$

$$\leftline{\indent$\displaystyle\eqalign{\phantom{I_2}
=\int {d^3k\over (2\pi)^3}\int_0^\infty dk_0 \bigg[
&{1\over p_0+\omega_k-\omega_{k+p}}\bigl (\delta(2k_0+p_0-\omega_k-\omega_{k+p}
+2\mu)\cr
&\phantom{1\over p_0+\omega_k-\omega_{k+p}}
-\delta(2k_0-p_0+\omega_k+\omega_{k+p}-2\mu)\bigr )\cr
&+p_\mu\rightarrow -p_\mu \biggr]\cr}$}$$

$$\leftline{\indent$\displaystyle\eqalign{\phantom{I_2}
={1\over 2}\int \,{d^3k\over (2\pi)^3}\,\bigl [
\, {1\over p_0+\omega_k-\omega_{k+p}}\bigl
(& \theta(\omega_k+\omega_{k+p}-p_0
-2\mu)\cr
&-\theta(p_0-\omega_k-\omega_{k+p}+2\mu)\bigr )
+p_\mu\rightarrow -p_\mu\bigr ] \cr}$}$$

$$\leftline{\indent$\displaystyle\phantom{I_2}=
-{1\over 2}\int {d^3k\over (2\pi)^3}\bigl [
{1\over p_0+\omega_k-\omega_{k+p}}\, {\rm sgn}
({p_0-\omega_k-\omega_{k+p}\over 2}+\mu)
+p_\mu\rightarrow -p_\mu\bigr ] .\hfil $\quad$ (27)$}$$
\noindent
With this correction, then, we obtain adding Eqs. (26) and (27)
$$\eqalignno{I_2+\tilde I_2={1\over 4}\int {d^3k\over (2\pi)^3}\bigl [
&{1\over p_0+\omega_k-\omega_{k+p}}\bigl ({\rm sgn}(\omega_k-\mu)
+{\rm sgn}(\omega_{k+p}
-p_0-\mu)\bigr )\cr
&+p_\mu\rightarrow -p_\mu\bigr ] .&(28)\cr}$$
\noindent
We recognize this to be exactly the same as Eq. (20) and it is clear that the
discrepancy lies truly in the combination of factors with opposite analytic
behaviour using the naive Feynman formula. The correct
 combination formula, first
derived by Weldon [7], is given in Eq. (8)
and, as is clear, is essential in problems even at zero temperature.
\bigskip
\bigskip
{\bf VII.  Conclusion :}
\bigskip
We have generalized Weldon's formula for combining products of factors and have
given an alternate derivation of this. This appears to be at the heart of the
disagreement about the analytic behaviour of Feynman amplitudes at finite
temperature. We have tried to emphasize with an example that the modified
Feynman parametrization formula is quite useful even at zero temperature. We
have systematically calculated the
 real part of the self-energy of a nonrelativistic, degenerate
electron gas at $T=0$ by direct integration of the delta function, by the
$\epsilon$-regularization (without Feynman parametrization, with naive Feynman
parametrization and with modified Feynman parametrization) and have shown that
it is not the $\epsilon$-regularization which is problematic. Rather, it is the
naive Feynman parametrization which needs to be modified according to Eq. (8).
In this case, all calculations (direct calculation, $\epsilon$-regularization
without Feynman parametrization as well as $\epsilon$-regularization
with modified Feynman parametrization) agree.

It is interesting to note that $Re \,\pi (p)$, even for the degenerate electron
 gas at $T=0$, has a nonanalytic behaviour
at $p^\mu=0$ (see for example ref. [9] pg 162).
 The Feynman rules at $T=0$
are, of course, fixed and cannot be changed {\it ad hoc}.
In fact, it is evident that whenever a physical system interacts with a
real background (such as the Fermi sea or the finite temperature thermal
bath), there would necessarily arise such nonanalyticity.  This appears to
be a general and physical effect and, therefore, we believe that modifying
the finite temperature Feynman rules to make the amplitude analytic is
inappropriate.  In fact, it would be quite interesting to study the
effective action for this system at
 zero or finite temperature in order to get a
deeper understanding of the additional Feynman rules at
$T \not= 0$ [10,11].

This work was supported in part by the U.S. Department of Energy Grant no.

\noindent DE-FG02-91ER40685 and P. B. acknowledges partial financial
support from CAPES.

\vfill
\eject
{\bf References:}
\bigskip
\item{1.} H. A. Weldon, Phys. Rev. {\bf D28}, 2007 (1983).
\item{2.} T. Matsubara, Prog. Theor. Physics {\bf 14}, 351 (1955).
\item{3.} P. S. Gribosky and B. R. Holstein,
 Z. Phys.  {\bf  C47}, 205 (1990).
\item{4.} H. Umezawa, H. Matsumoto and M. Tachiki, \lq\lq Thermo
Field Dynamics and Condensed States", North-Holland,
 Amsterdam, 1982.
\item{5.} Y. Fujimoto and H. Yamada, Z. Phys. {\bf C37}, 265 (1988).
\item{6.} P. F. Bedaque and A. Das, Phys. Rev. {\bf D45}, 2906 (1992).
\item{7.} H. A. Weldon, West Virginia University
 Preprint WVU-992 (August, 1992).
\item{8.} A. A. Abrikosov, L. P. Gorkov and I. E. Dzyaloshinski,
\lq\lq Methods of Quantum Field Theory in Statistical Physics", Dover,
New York 1963.
\item{9.} A. L. Fetter and J. D. Walecka,
 \lq\lq Quantum Theory of Many Particle Systems", McGraw-Hill,
New York 1971.
\item{10.} H. Matsumoto, Y. Nakano and H. Umezawa, Phys. Rev. {\bf D31},
 1495 (1985).
\item{   } T. S. Evans, Z. Phys. {\bf C36}, 153 (1987).
\item{11.} R. L. Kobes, Phys. Rev. {\bf D43}, 1269 (1991).
\item{   }  T. S. Evans,
Nucl. Phys. {\bf B374}, 340 (1992).
\bye